\documentclass[traditabstract]{aa} 
\usepackage{graphicx}
\usepackage{txfonts}
\usepackage{natbib}
\begin{document}
   \title{Characterization of Solar Telescope Polarization Properties Across
   the Visible and Near-Infrared Spectrum}

   \subtitle{Case Study: The Dunn Solar Telescope}

   \author{H. Socas-Navarro\inst{1,2}
   \and
   D. Elmore\inst{3}
   \and
   A. Asensio Ramos\inst{1,2}
   }

   \institute{Instituto de Astrof\'\i sica de Canarias,
     Avda V\'\i a L\'actea S/N, La Laguna 38200, Tenerife, Spain
   \and
   Departamento de Astrof\'\i sica, Universidad de La Laguna, 38205, 
   La Laguna, Tenerife, Spain 
   \and
   National Solar Observatory/Sacramento Peak,
   3010 Coronal Loop, Sunspot, NM 88349, USA
 }

   \date{Received September 15, 1996; accepted March 16, 1997}

   \authorrunning{Socas-Navarro et al}
   \titlerunning{Solar Telescope Polarization Characterization}

\abstract
{
Accurate astrophysical polarimetry requires a proper characterization
of the polarization properties of the telescope and instrumentation
employed to obtain the observations. Determining the telescope and
instrument Muller matrix is becoming increasingly difficult with the
increase in aperture size of the new and upcoming solar telescopes. We
have carried out a detailed multi-wavelength characterization of the
Dunn Solar Telescope (DST) at the National Solar
Observatory/Sacramento Peak as a case study
and explore various possibilites for the determination of its
polarimetric properties. We show that the telescope model proposed in
this paper is more suitable than that in previous work in that it
describes better the wavelength dependence of aluminum-coated
mirrors. We explore the adequacy of the degrees of freedom allowed by
the model using a novel mathematical formalism. Finally, we
investigate the use of polarimeter calibration data taken at different
times of the day to characterize the telescope and find that very
valuable information on the telescope properties can be obtained in
this manner. The results are also consistent with the entrance window
polarizer measurements, opening very interesting possibilities for the
calibration of future large-aperture solar telescopes such as the ATST
or the EST.
}


   \maketitle
%

\section{Introduction}

The observation of polarization in astrophysical objects allows us to
measure magnetic fields in their environment or to learn about the
physical conditions reigning in the regions where light is scattered
into our line of sight. However, polarimetry is a very challenging
technique because the signals to measure are typically very weak ($<
1\%$ of the observed intensity) and because the telescope and
instrumentation employed introduce spurious polarization. Most
polarimeters have calibration optics to determine its polarimetric
properties so that it is possible to remove the instrumental
contamination from the observed signals. However, it is only possible
to characterize the instrumentation downstream from the calibration
optics. Ideally, one would like then to have calibration polarizers
before the telescope primary mirror and covering the entire
aperture. Unfortunately, this is impractical in most
situations. Currently, only the Dunn Solar Telescope at the National
Solar Observatory/Sacramento Peak Observatory (Sunspot, NM, USA), the
German VTT at the Observatorio del Teide on the island of Tenerife and
the Swedish Solar Telescope at the Observatorio del Roque de los
Muchachos on the island of La Palma (both operated by the Instituto de
Astrof\'\i sica de Canarias, Spain) have that capability
\citep{SLMP+97, BSC+05, S05}. Even in these three cases the operation of the
telescope calibration devices is far from routine and it takes
considerable effort and a full day (sometimes more) of continued
observation.

The largest solar telescopes currently under operation do not exceed
1~m of aperture but the soon-to-be commissioned Gregor has 1.5~m
\citep{VvdLK+07} and plans already exist for the construction of two
4~m telescopes: the Advanced Technology Solar Telescope (ATST,
\citealp{KRH+02}) and the European Solar Telescope (EST,
\citealp{EST10}). With such large apertures, full telescope
calibrations become extremelly challenging from a technical
standpoint.

An additional problem is that the configuration of the telescope is
not fixed. It has some degrees of freedom, e.g. to be able to point at
different coordinates on the sky. When the telescope moves, the angles
among some of the many mirrors and optical elements along the light
path also change. In the case of solar observations there is typically
a continuous variation of at least two mirrors as one tracks the
apparent motion of the Sun on the sky. Therefore, it does not suffice
to derive the Muller matrix at a given time. We need to know how it
depends on the telescope configuration. In this manner, since we know
the specific configuration at the time of each observation, we can use
the correct Muller matrix to calculate the parasitic instrumental
polarization induced and remove it from the data.

We shall follow here a similar nomenclature to that of \citet{SLMP+97}
and \citet{SNEP+06}. We break down the polarimetric measurement process as:
\begin{equation}
\vec S_{meas} = \tens{X} \tens{T}(\vec \alpha) \vec S_{\sun} \,
\end{equation}
where $\tens{T}(\vec \alpha)$ is the telescope Muller matrix,
$\tens{X}$ denotes the polarimeter response, and $\vec S_{\sun}$ and
$\vec S_{meas}$ are the incoming (solar) and the measured Stokes
vectors, respectively. The polarimeter calibration optics (typically a
combination of a polarizer and a retarder than can be slided in and
out of the beam and rotated independently of each other) mark the
split point of the optical train. Any optical surface upstream from
that point is considered part of the telescope and included in
$\tens{T}$, whereas everything downstream is part of the polarimeter
and characterized in $\tens{X}$.  We shall take the polarimeter as a
static system since it has no moving parts, with only minor changes
due to thermal fluctuations. The telescope, on the other hand, has a
variable configuration, e.g. with moving mirrors to point and track
across the sky. We parameterize the particular configuration in the
vector $\alpha$.

Acquiring calibration data to constrain $\tens{T}(\vec \alpha)$ is
much more difficult and time-consuming than for $\tens{X}$. This is
due to two reasons: a)the fact that $\alpha$ (and therefore
$\tens{T}$) varies over the course of a day, and b)the solar beam has
a much larger diameter at the entrance of the telescope than at the
polarimeter. The first difficulty imposes the need to take calibration
observations for at least a half day (but preferably more than that)
to ensure appropriate coverage over the range of variation of the
parameters in $\vec \alpha$. The second problem is not insurmountable
for currently existing 1~m aperture telescopes but the planend large
aperture of the EST or the ATST will require new strategies (e.g.,
\citealp{SN05a, SN05b}).

In this paper we take the DST as a case study and analyze its
polarimetric properties at many wavelengths spanning the visible and
near-infrared (nIR) ranges of the spectrum. We start by building an
improved model of the telescope with respect to what has been done in
previous work. We then use a novel mathematical formalism to validate
the degrees of freedom in the model. Finally, we use two different
strategies to fit the various parameters and obtain a reliable
multi-wavelength characterization of the telescope. One of such
strategies makes use of data taken with entrance window polarizers in
the beam, whereas the other uses solar data thus avoiding the need for
polarizers filling the full telescope aperture. We conclude that both
strategies produce consistent results, which opens new interesting
perspectives for the calibration of future large-aperture facilities.

\section{The telescope model}
\label{sec:telescopemodel}

In observing mode, the DST has the following optical surfaces, which
could in principle alter the polarization state of the solar light. In the
order encountred by the incoming beam, we find:
\begin{itemize}
\item An entrance window (EW) used to keep the optical train
  evacuated. Mechanical stress on the window mount could make it act
  as a retarder with a small degree of retardation.
\item A turret with two 1~m diameter mirrors that track the Sun and
  send the light down in the vertical direction to the primary mirror
  which is located underground. The first turret mirror moves in the
  elevation direction ($\gamma$) and the second in azimuth
  ($\phi$). These two angles are needed to define the telescope
  configuration and we take them as the first components of the
  configuration vector $\vec \alpha$ introduced earlier. The turret
  is a heavily polarizing device, since the beam strikes both mirrors
  at a 45~degree angle of incidence.
\item The primary mirror, which due to its near normal angle of
  incidence does not alter the light polarization significantly except
  for a 180~degree phase change.
\item The exit window (XW), which marks the end of the evacuated optical
  train. Like the EW, this element could introduce some small degree
  of retardation in the beam (in general, different from that of the
  EW). The main mirror, the XW and the instrument platform can rotate
  rigidly to compensate for the diurnal solar image rotation on the
  instrument focal plane and/or to define the orientation of the
  spectrograph slit. Let us donte by $\psi$ the angle of this whole
  system, which is the third and last element of the configuration
  vector $\vec \alpha$.
\end{itemize}

Behind the XW we have the polarimeter calibration optics and the
polarimeter itself. Therefore, the above elements are all that we
need to consider in our telescope model. 

We model both windows as an ideal retarder whose retardation is a free
parameter. The orientation of the retarder fast axis is also a free
parameter. The turret mirrors are modeled taken their
diattenuation ($r_s/r_p$) and retardance as free parameters and
calculating the orientation of the plane of incidence from $\gamma$
and $\phi$. For the main mirror it is a good approximation to consider
a perfectly symmetric reflection with no diattenuation and a
180-degree retardation. With these considerations in mind, we
construct the total Muller matrix of the telescope as:

\begin{equation}
\label{eq:T}
\tens T(\gamma, \phi, \psi)=\tens D_{XW} \tens M_{Main} \tens R_{Main-AZ}(\psi,\phi) \tens
M_{AZ} \tens R_{EL}(\gamma) \tens M_{EL} \tens D_{EW}  \, ,
\end{equation}
where $\tens D$ denotes the Muller matrix of a retarder in the ($s,p$)
reference frame (i.e., with the axes parallel and perpendicular to the
incidence plane), $\tens M$ is the matrix of a mirror and $\tens R$ is
a rotation of the coordinate frame from one element to the next. The
subscripts $EL$, $AZ$ and $Main$ refer to the elevation and azimuth
mirrors of the turret and the primary mirror, respectively. In the
equation above we have only written down explicitly the dependence of
the various matrices with the telescope configuration angles $\vec
\alpha$, but not with the free parameters.

The free parameters of the model are then the EW fast axis orientation
and retardance, the elevation and azimuth mirrors diattenuation and
retardance and the XW fast axis orientation and retardance. In addition
to those six parameters, we also consider as a free parameter a
rotation angle between the telescope and the polarimeter respective
reference frames and finally, in the case that the entrance window
calibration polarizer is used, the zero point of the calibration
polarizer. This results in a total of 8 free parameters for a
single-wavelength model. In the next section we present a formal
justification that this number of free parameters is nearly optimal
for the problem under consideration.

For a multi-wavelength characterization we take a somewhat different
approach from that in \citet{SNEP+06}. We have observed that the
polynomial fit proposed in that work to the wavelength dependence of
the various parameters is not always adequate, as it does not always
capture the real polarimetric behavior of the optical elements. When
the number of wavelengths observed increases, that model has
difficulty fitting all the data. In view of the results presented in
this paper, particularly those in section~\ref{sec:EWpol} below, it is
easy to see that a third-order polynomial will not be able to
reproduce the real behavior of the telescope at all wavelengths.

The new model that we propose in this work has a number of $4+4\times
n_{\lambda}$ free parameters (where $n_{\lambda}$ is the number of
wavelengths observed). The 4 wavelength-independent parameters are the
EW and XW fast axis orientations, the telescope-polarimeter reference
frame rotation and the offset of the EW polarizers with respect to our
assumed zero point. The EW, XW retardances and the turret mirror
diattenuations ($r_s/r_p$) and retardances are functions of
wavelength. We take their value at the observed wavelengths as a free
parameter. Intermediate values are obtained from linear
interpolation. In this manner increasing $n_{\lambda}$ results in more
free parameters but at the same time the amount of data is also
largely increased.

\section{Dimension analysis}

In principle, even if one bases the model of the telescope on simple
assumptions, it is possible that the final model contains too many
free parameters that cannot be constrained by the observations.  In
such a case, when one fits the model parameters to a set of
calibration observations, the model might not be representative of the
general behavior of the telescope. Obviously, this is produced by the
overfitting ability of a model with too many free parameters. This is
particularly relevant when several parameters are degenerated, meaning
that the variation of one parameter can be compensated to a great
extent with variations in one or more of the other parameters.

Consequently, we analyze the intrinsic dimensionality of the model
using the maximum-likelihood estimation developed by
\cite{levina_bickel05} and applied with success by
\cite{asensio_dimension07} to estimate the intrinsic dimensionality of
spectro-polarimetric data. By intrinsic dimensionality we mean the
number of free parameters that the $\tens{T}$-matrix really
depends on, taking into account that degeneracies introduce
correlations between the parameters and reduce the
dimensionality. Given $N$ vectors of dimension $M$ represented as
$\vec{x}_i$, the dimensionality is estimated by using the
expression:
\begin{equation} \hat{m}_k^{-1} = \frac{1}{N (k-1)} \sum_{i=1}^N
\sum_{j=1}^{k-1} \log \frac{T_k(\mathbf{x}_i)}{T_j(\mathbf{x}_i)}.
  \label{eq:average2}
\end{equation} where $T_k(\mathbf{x}_i)$ represents the Euclidean
distance between point $\mathbf{x}_i$ and its $k$-th nearest
neighbor. The previous equation is only valid for $k>2$ and it depends
on the number of neighbors that we select. In principle, this can be
used to analyze variations of the intrinsic dimensionality at
different scales, but our results are relatively constant with
$k$. The computational cost of this method is mainly dominated by the
calculation of the $k$ nearest neighbors for every point
$\mathbf{x}_i$.

\begin{figure}
  \centering
  \includegraphics[width=9cm]{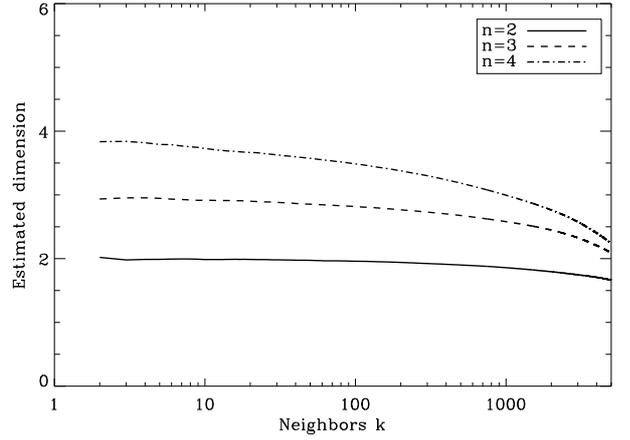}
  \caption{Dimensionality of three different polynomial models with 2,
  3 and 4 coefficients. The curves converge to the correct degrees of
  freedom for a small number of neighbors.}
  \label{fig:polynomial}
\end{figure}

As an illustrative example, we have considered data generated with a
polynomial function:
\begin{equation}
  y(x)=\sum_{i=0}^{n-1} c_i x^i \, .
\end{equation}
This function may be viewed as a non-linear model with $n$ free
parameters (the $c_i$ coefficients). Our aim is to estimate the order
of the polynomial just from the samples. Since we have generated the
data for this experimient, we can then verify {\em a posteriori} that
the results accurately yield the correct number. Three different
experiments were carried out for polynomials of order 1, 2 and 3,
respectively. For each value of $n$, we generate $N=10^4$ vectors
composed of samples of the polynomial at $M=10$ different positions
($x$). The estimation of the dimensionality is shown in
Fig~\ref{fig:polynomial} where $n$ indicates the number of
coefficientes of the polynomial (i.e., the polynomial order is
$n-1$). Note that the results converge towards the correct
dimensionality for small number of neighbors (for large values, the
results are sensitive to the finite and discrete nature of the
grid). Further details of this procedure and more exhaustive tests can
be found in \cite{asensio_dimension07}. Here we simply intend to use
this example to illustrate the application to the telescope model
presented below, where instead of a simple polynomial we have the
$\tens{T}$-matrix constructed as indicated in Eq~\ref{eq:T} above from
its 8 free parameters. If we had correlations or degeneracies among
these parameters, then the dimensionality of the data produced with
the model would be less than the number of free parameters.

For the analysis of the telescope model we consider each Stokes
parameter $Q$, $U$ and $V$ separately, and the $N$ vectors are built
as follows.  Let $N_\mathrm{pol}$ be the number of angles of the axis
of our EW polarizers. Let $N_\mathrm{ang}$ be the number of
combinations of azimuth, elevation and table angles that 
characterize the telescope configuration. For each combination of
polarizer angle, azimuth, elevation and table angle, we propagate a
Stokes vector representing unpolarized light through the telescope
(with its EW polarizers) by multiplying $(1,0,0,0)\dagger$ by the full
telescope Muller matrix $\tens{T}$ (the symbol $\dagger$ represents
the matrix transposition operation).  Keeping the parameters of
the matrix fixed, we construct the vector of length $M=N_\mathrm{pol}
N_\mathrm{ang}=100$ by stacking the emergent Stokes parameter ($Q$,
$U$ or $V$) for all the possible combinations. Each such vector then
represents a realization of the observable that can be used to
characterize the Mueller matrix of the telescope.  This procedure is
repeated $N$ times until the entire database is filled.  Due to
computational limitations in the $k$ nearest neighbors calculation, we
limit ourselves to $N=10^4$ different values of the parameters. These
values have been generated by means of a latin hypercube sampling \citep{MKBC79},
which produces a better sampling of the parameter space.

\begin{figure}
  \centering
  \includegraphics[width=9cm]{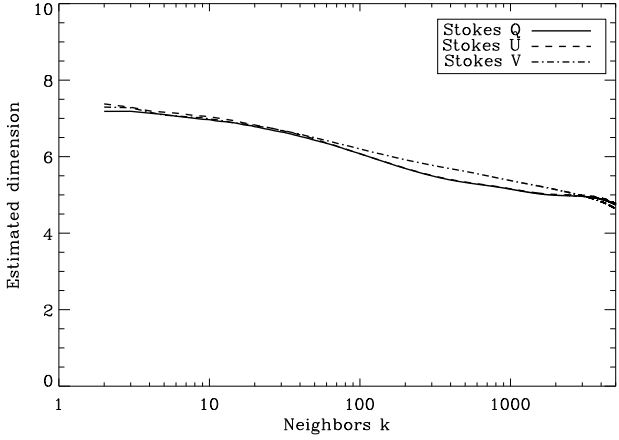}
  \caption{Dimensionality of the telescope model with 8 free
    parameters described in Section~\ref{sec:telescopemodel}. All
    Stokes parameters $Q$, $U$ and $V$ converge to a value of
    approximately 7.5, evidencing that the model does not have
    degerate parameters.}
  \label{fig:dimensionality8}
\end{figure}

\begin{figure}
  \centering
  \includegraphics[width=9cm]{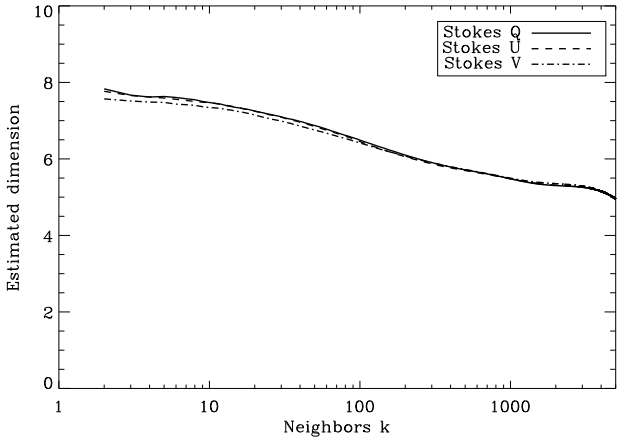}
  \caption{Dimensionality of the telescope model with 10 free
    parameters (the previous 8 plus main mirror diattenuation and
    retardance). All Stokes parameters $Q$, $U$, and $V$ converge to a
  value of approximately 8 even though we have 10 free parameters,
  evidencing that the model has degenerate parameters.}
  \label{fig:dimensionality10}
\end{figure}

We have applied the dimension analysis on these data and obtained the
results plotted in Fig~\ref{fig:dimensionality8}. All of the Stokes
parameters exhibit the same behavior and converge to approximately
7.5, which is very close to the number of free parameters (8, see
Section~\ref{sec:telescopemodel}) in our model. From this we can
conclude that no significant degeneracies exist among the various free
parameters and that a variation on each one of them produces an
independent, measureable result on the observables. In other words, we
can be confident that with enough data and a sufficient coverage of
the configuration space, it is possible to univocally retrieve all of
these parameters.

Our original model also had the main mirror diattenuation and
retardance as free parameters. However, after a few attempts with
different initializations, we quickly realized that there were
uniqueness issues as we were able to fit the data with different
combinations of the parameters. In particular, we found that the main
mirror retardation exhibited a seemingly random wavelength dependence
that was nearly identical (but opposite) to that of the XW. A quick
look at the model (see Eq~\ref{eq:T}) shows that there are no other
elements between the main mirror and the XW. Therefore, one can set
any arbitrary value for the retardation in the main mirror and then
compensate it with an opposite retardation in the XW. We explored this
issue with the dimension analysis, this time having 10 free parameters
in the model (the previous 8 plus the main mirror diattenuation and
retardance). The results are plotted in
Fig~\ref{fig:dimensionality10}. Note that, even though we now have
more free parameters, the dimensionality of the data has not changed
significantly and we obtain again a result close to 8. This indicates
that this model now has too much freedom and some of the free
parameters are degerate. We thus decided to fix the main mirror
properties to those of a non-polarizing reflection, which is a good
approximation anyway based on symmetry considerations.

\section{Entrance window polarizers}
\label{sec:EWpol}

The DST is equipped with an array of achromatic linear polarizers that
can be mounted on top of the EW. The entire array may be rotated in
azimuth to any desired angle by means of a system consisting of a
motor and its associated control electronics. With this device it is
possible to feed the telescope with light in a known state of
polarization and probe the properties of the full optical train, from
the EW to the polarimeter. 

In addition to the EW polarizers we also have the regular polarimeter
calibration optics with which it is possible to fully characterize the
instrument (in our case, SPINOR). We start the process by determining
the SPINOR response matrix which we then fix in the determination of
the telescope properties. This is a routine operation that involves
inserting the SPINOR calibration polarizer and retarder and rotating
them independently to various angles. After going through the
calibration polarizer, the previous state of polarization becomes
irrelevant as the light will then become fully polarized in the
direction set by the polarizer (the total light intensity is also
irrelevant since we work with normalized Stokes vectors and consider
only the degree of linear/circular polarization). The polarimeter
calibration is then independent of the telescope configuration
($\alpha$).

\begin{figure}
  \centering
  \includegraphics[width=9cm]{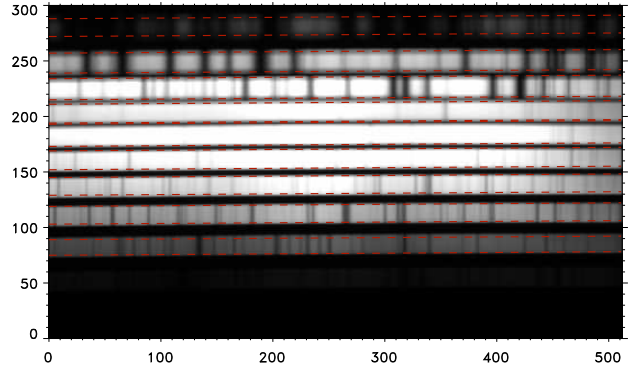}
  \caption{ Example of a calibration image with the cross-dispersed
    spectral orders. Each band enclosed with dashed lines represents
    one of the overlapping orders in the spectrograph (order-isolation
    filters are removed for these operations) that we have used for
    this work, ranging from 470 (bottom) to 1413 (top) nm. Spectral
    lines are visible in the data. The central orders have been
    saturated in the figure to show the weaker ones at the top and the
    bottom.  }
  \label{fig:spectra}
\end{figure}

It is important to have the polarimeter characterized first,
othwerwise we would have to fit also the matrix $\tens{X}$ and there
would be too many free parameters with unpleasant couplings between
some of the optical elements. On 2010 May 3 we performed a total of
21 polarimeter calibration operations at different times during the
day. In each one of these operations we recorded a sequence of
76$\times$8 images (76 configurations of the calibration optics and 8
modulation states) with a cross-dispersing prism placed in front of
the detector, removing the order isolation pre-filter and blocking
most of the spectrograph slit length to allow only a small field of
view in the spatial direction. After proper demodulation, the 8 images
in the modulation sequence are transformed into 4 containing at each
pixel the Stokes $I$, $Q$, $U$ and $V$ parameters. Each image contains
a number of spectral ranges (9 in our case) that span the entire
visible and nIR range, as shown in the example of
Fig~\ref{fig:spectra}.

We employ a Levenberg-Marquardt (see, e.g. \citealt{PFT86}) algorithm
to fit the Stokes data to a model with a 4$\times$4 response
matrix. Since the calibration retarder is not a perfect $\lambda$/4
plate over the entire wavelength range, we also take its retardance as
a free parameter and determine it from the fit. The orientation of the
retarder fast axis is also a free parameter to correct for possible
errors in the mount alignment. All the Stokes vectors are normalized
to their respective intensity so only their orientation in the
Poincar\' e sphere is considered. As a result, the $\tens{X}_{(1,1)}$
matrix element will always be equal to 1.

\begin{figure}
  \centering
  \includegraphics[width=8cm]{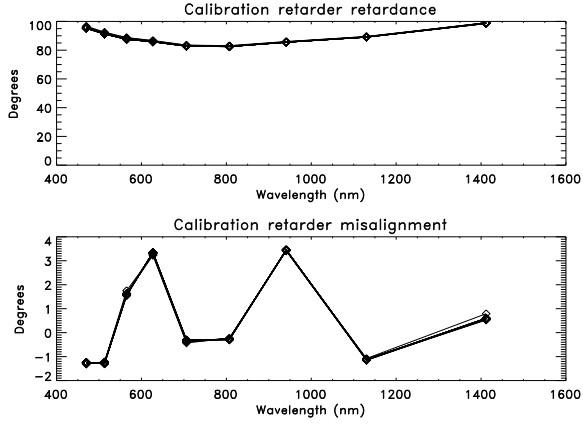}
  \caption{Properties of the polarimeter (SPINOR) calibration retarder
    as a function of wavelength. Upper panel: Retardance. Lower panel:
    Difference between the retarder fast axis orientation and the
    mount zero point. In both cases we have overplotted all 21 curves
    obtained from the (independent) calibration measurements carried
    out over the course of a day.}
  \label{fig:xmatrix1}
\end{figure}

\begin{figure*}
  \centering
  \includegraphics{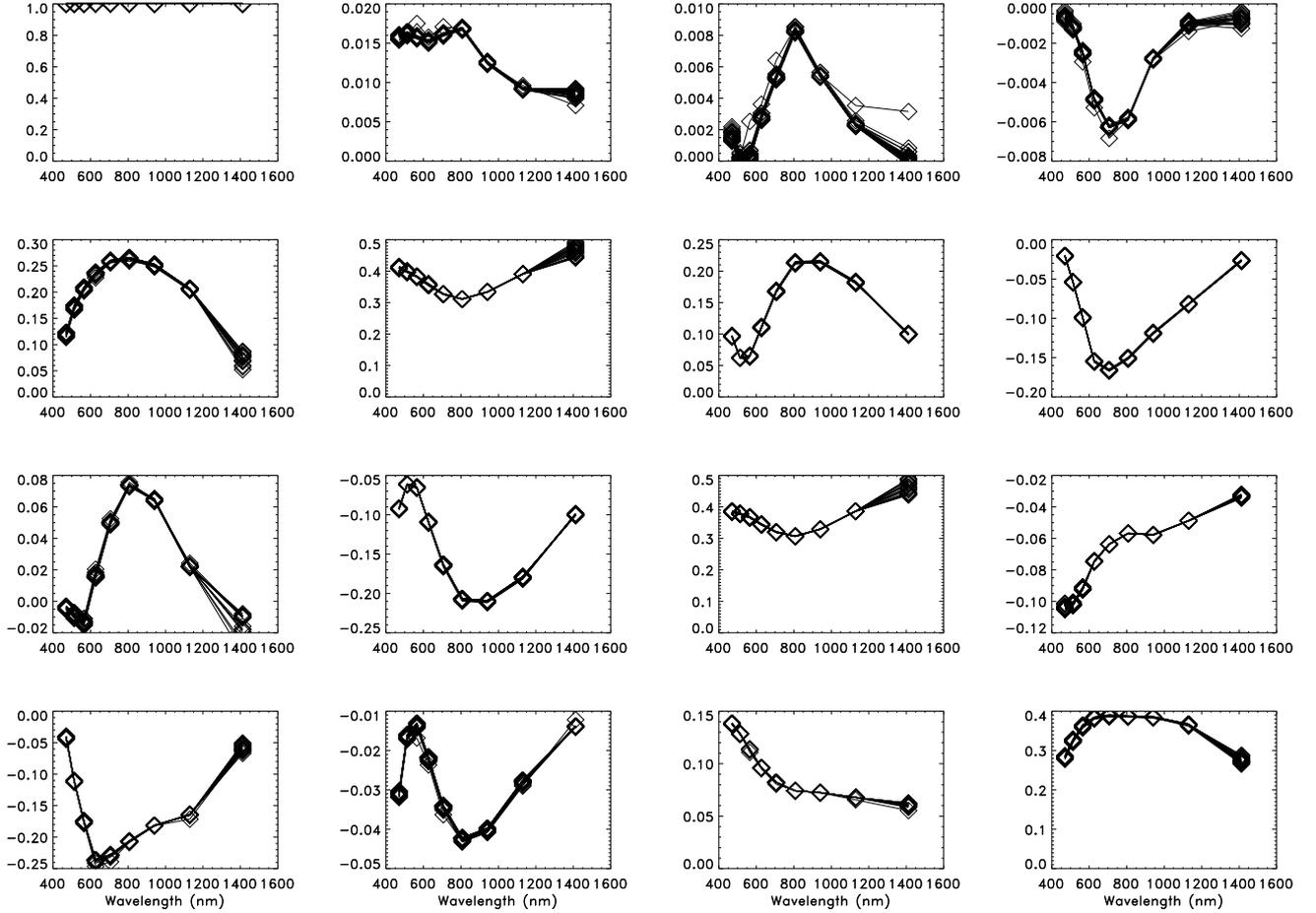}
  \caption{Elements of the polarimeter (SPINOR) 4$\times$4 response
    matrix as a function of wavelength. In all panels we have
    overplotted all 21 curves obtained from the (independent)
    calibration measurements carried out over the course of a day.}
  \label{fig:xmatrix2}
\end{figure*}

Figures~\ref{fig:xmatrix1} and~\ref{fig:xmatrix2} show the resulting
polarimeter properties as a function of wavelength obtained as
described above. In Fig~\ref{fig:xmatrix1} we can see
the properties of the calibration retarder. Since the retarder is not
perfectly achromatic, there is a variation of its retardance
(Fig~\ref{fig:xmatrix1}, upper panel). The difference between the
orientation of the retarder fast axis and its reference zero-point is
also fitted (Fig~\ref{fig:xmatrix1}, lower panel). As expected this
difference is very small, below a few degrees in any case.

Figure~\ref{fig:xmatrix2} shows the 16 elements of the $\tens{X}$
matrix as a function of wavelength. As mentioned above, we have
repeated the measurements 21 times at different times of the day. Both
figures are actually showing all 21 curves overplotted. The
differences among them are so small in most cases that all of these
curves virtually coincide (although the spread seems to increase for
the greatest wavelengths). This impressive agreement reinforces our
degree of confidence in the methodology that we have employed, since
each one of the 21 curves was obtained from independent measurements
that were also fitted independently. Furthermore, it also indicates
that SPINOR exhibits a very high degree of temporal stability in its
polarimetric properties.

Now that we have the elements of $\tens{X}$ and we can fix that part
of the equation, we turn to the telescope itself. With the EW
polarizer in the beam, we acquired data during the afternoon of 2010 May
3 and also during the following day. A total of 15070 Stokes
vectors were recorded at each one of the 9 wavelengths considered (the
same wavelengths that had been observed before during the polarimeter
calibration) and for different telescope configurations, which was
continuously tracking the Sun on the sky and also moving the DST
rotating platform to different angles. 

\begin{figure*}
  \centering
  \includegraphics{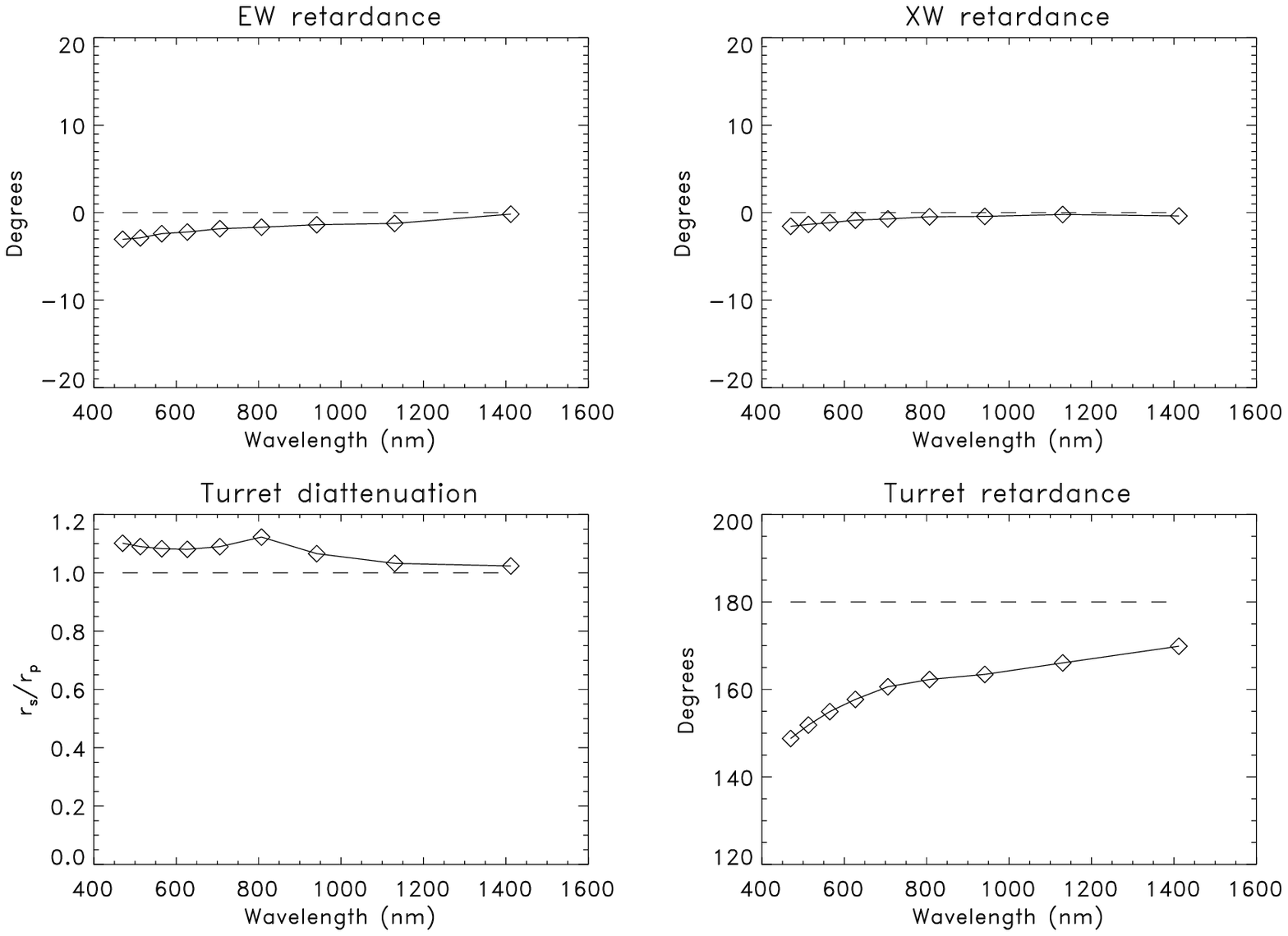}
  \caption{Wavelength-dependent parameters obtained for the telescope
    model. The dashed lines represent the properties of a
    non-polarizing element, such as the DST primary mirror in our
    model.}
  \label{fig:tmat_wave}
\end{figure*}

Similarly to the polarimeter characterization above, we applied a
computer-intensive Levenberg-Marquard fit to the entire dataset using
the telescope model described in Section~\ref{sec:telescopemodel}. The
results are summarized in Fig~\ref{fig:tmat_wave}
(wavelength-dependent parameters) and in Table~\ref{table:tmat}
(wavelengt-independent parameters). Comparing Fig~\ref{fig:tmat_wave}
to Fig~5 of \citet{SNEP+06} one can see where the problems with the
previous model come from. The third-order polynomials can adequately
reproduce the behavior that we find here for the EW and XW retardances
and also the turret retardance. However, the turret $r_s/r_p$ is
not properly described and, for some wavelengths, it departs significantly.

\begin{table}
  \caption[]{Telescope wavelength-independent parameters from the model fit.}
  \label{table:tmat}
  $$ 
  \begin{array}{p{0.5\linewidth}l}
    \hline
    \noalign{\smallskip}
    Parameter      &  Value \\
    &     (degrees) \\
    \noalign{\smallskip}
    \hline
    \noalign{\smallskip}
    EW fast axis orientation &  138.29  \\
    XW fast axis orientation &  42.76  \\
    Telescope-SPINOR frame rotation & 93.17 \\
    EW polarizer zero offset & 81.53 \\
    \noalign{\smallskip}
    \hline
  \end{array}
  $$ 
\end{table}

\begin{figure}
  \centering
  \includegraphics[width=9cm]{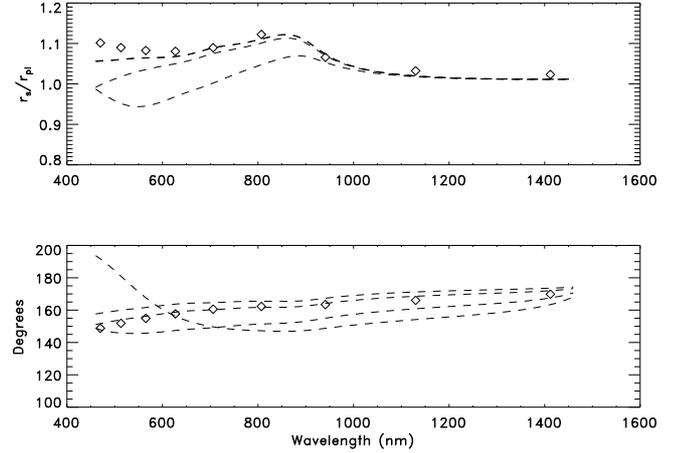}
  \caption{Theoretical calculation of an aluminum mirror
    diattenuation (upper panel) and retardance (lower panel) as a
    function of wavelength. The four curves represent different
    values for the thickness of the oxide layer. From top to bottom:
    10, 20, 50 and 80~nm, respectively. The curves for 10 and 20~nm
    overlap in the upper panel. The diamonds show the values obtained
    from our fit for the turret mirrors (see Fig~\ref{fig:tmat_wave})
    }
  \label{fig:zemax}
\end{figure}

We can see in Fig~\ref{fig:tmat_wave} that the various elements behave
monotonically, with a trend for the instrumental polarization to
decrease towards longer wavelengths (note, however, that even in the
nIR the telescope elements polarize significantly). The only exception
is the diattenuation of the turret mirrors, which exhibits a peak
around 850~nm. This peak is to be expected for an aluminum-coated
mirror. Theoretical models of the mirrors show a qualitatively similar
behavior with the 850~nm peak. The actual details depend on the
thickness of the Al$_2$O$_3$ layer deposited on the mirror substrate
but some ilustrative examples are given in Fig~\ref{fig:zemax}. The
details of the calculation can be found in \citet{BW75}.

\section{Polarimeter calibration optics}
\label{sec:davidcal}

When incoming unpolarized light goes through the telescope system, it
becomes partly polarized. The state of polarization depends on the
telescope configuration $\alpha$. It is then possible to obtain
information on the telescope properties by simply monitoring how the
transfer from Stokes~$I$ to $Q$, $U$ and $V$ changes over the course
of the day. Such measurement can in principle be carried out without
resorting on polarizers filling the entire telescope aperture, as done
in Section~\ref{sec:EWpol}. We can use the polarimeter calibration
optics at the exit port of the telescope to measure the outgoing
Stokes vector produced from a raw unpolarized solar beam.

The main polarization creation device is the turret, which
introduces both diattenuation and retardation in the unpolarized
beam. The resulting partly polarized beam further undergoes an
additional retardation by the XW. The main mirror contributes
negligibly, as mentioned above, because of the near normal
incidence. Finally, the retardance introduced by the EW on an
unpolarized incoming beam is also irrelevant. Based on these
considerations, it is easy to see that this method will not provide
information on the EW or the main mirror properties but one may hope
to learn something about all the other elements.

\begin{figure*}
  \centering
  \includegraphics{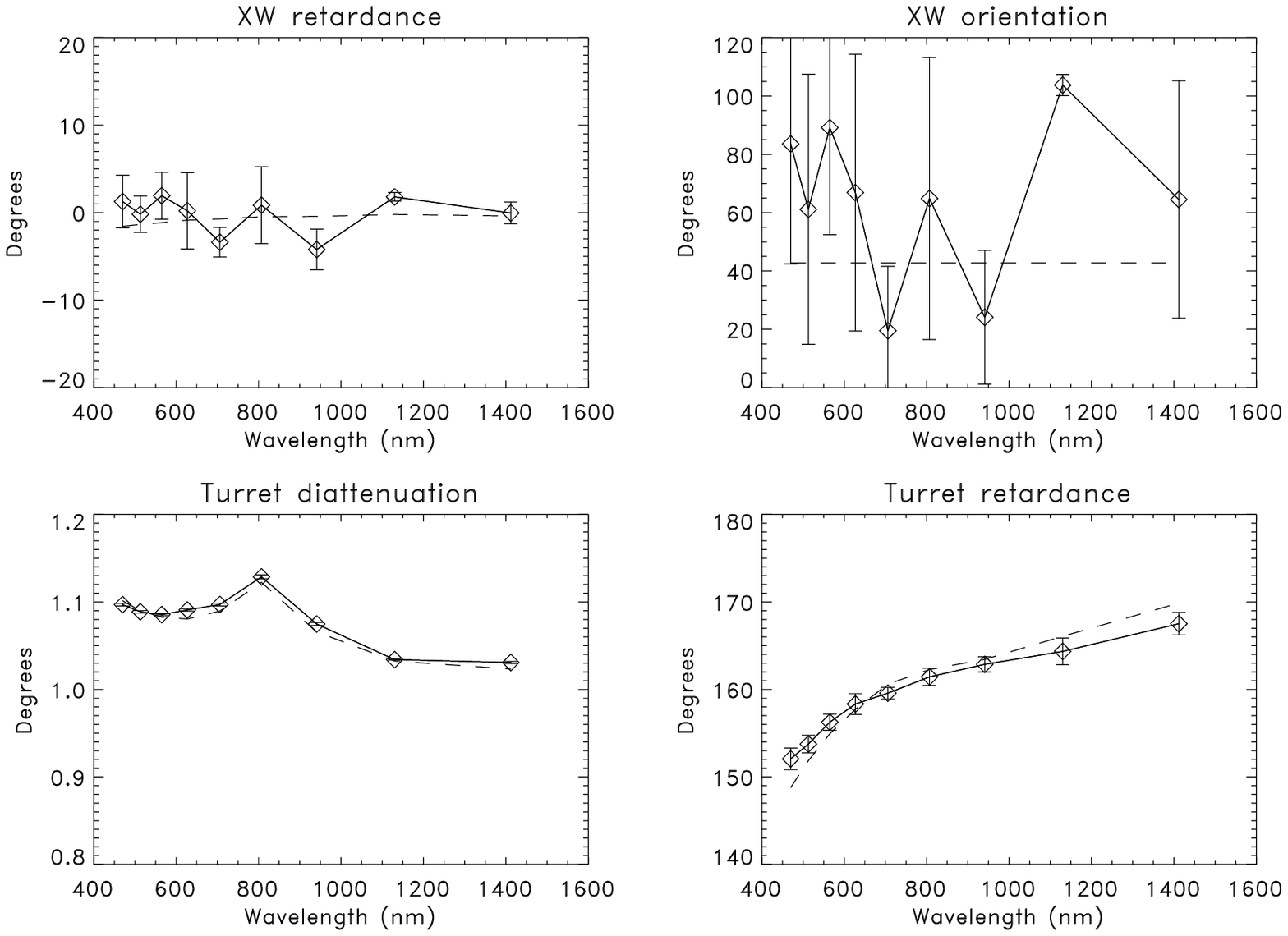}
  \caption{Fit to the data with the polarimeter calibration optics
    (diamonds) compared to the model obtained with the EW polarizer
    (dashed line). 
    }
  \label{fig:elmore}
\end{figure*}

Figure~\ref{fig:elmore} shows the results of fitting all the
calibration operation data acquired over the course of a day to the
telescope model. The results are compatible with the measurements
using the EW polarizer presented in Section~\ref{sec:EWpol}. The
turretm mirror properties are much better constrained than those of
the EW, as one would expect since they polarize the incoming beam much
more strongly than the XW.

\section{Conclusions}

Calibrating the instrumental polarization of large solar telescopes is
going to be an important challenge in the near future, especially for
multi-wavelength observations. A key part of the process is the
parametrization of the system in terms of a geometrical model with a
few free parameters which are determined by fitting large calibration
datasets. One needs to make sure that the model chosen has the right
number of free parameters. With too much freedom one is able to fit the data
but the model obtained is not unique and the properties of the
individual components are unreliable. Too little freedom, on the other
hand, limits the ability of the model to fit the calibration data and
results in an inaccurate calibration. We have presented here a robust
model for the DST telescope. Our dimension analysis, together with the
model's ability to fit all the data at all wavelengths, shows that it
has the correct amount of freedom. 

The technique based on EW polarizers is the most straightforward and
accurate way to characterize the polarimetric properties of a
telescope. However, we have shown that, when this is not practical, it
is also possible to use the calibration optics downstream to constrain
the model parameters (at least those related to the most heavily
polarizing elements) if a sufficiently large collection of calibration
data is acquired. These results are encouraging and open new
possibilities for accurate broadband characterization of future
large-aperture telescopes.

\begin{acknowledgements}
  The DST calibration data have been collected by the DST observing
  staff: Doug Gilliam, Mike Bradford and Joe Elrod. The National Solar
  Observatory is operated by the Association of Universities for
  Research in Astronomy under sponsorship of the National Science
  Foundation. HSN and AAR gratefully acknowledge financial support by
  the Spanish Ministry of Science and Innovation through project
  AYA2007-63881 (Solar Magnetism and High-Precision
  Spectropolarimetry).
\end{acknowledgements}

\bibliographystyle{aa.bst}

\bibliography{../bib/aamnem99,../bib/articulos}


\end{document}